# DHX36-mediated G-quadruplex unfolding is ATP-independent?

Hai-Lei Guo, Wei-Fei Chen, Stephane Rety, Na-Nv Liu, Ze-Yu Song, Yan-Xue Dai, Xi-Miao Hou, Shuo-Xing Dou & Xu-Guang Xi

Recently, Chen *et al.*[1] solved the crystal structure of bovine DHX36 bound to a G-quadruplex (G4) linked with a single-stranded DNA segment, revealing the details of interactions between the two molecules. By combining the crystal structure with single-molecule fluorescence resonance energy transfer (smFRET) analysis, they proposed that ATP-independent structural changes of the helicase remodel G4, resulting in a substrate unwound by a single nucleotide; furthermore, the G4 may be unfolded repetitively by DHX36 without ATP. The model and conclusion are very interesting. Here, however, we provide evidences from several different assays that may challenge them.

We noticed two details in their protein-DNA complex preparation and crystallization protocol[1] that might cause problems. First, the protein without glycine-rich-element, DHX36-DSM, was modified with mutation of 6 amino acids and dimethylation of 63 lysines. It is not very clear if such treatments have disturbed the intermolecular interactions in their protein-DNA complex or not. Second, the linear DNA substrate (DNA$^{Myc}$) that harbors a *Myc*-promoter-derived G4-forming sequence was directly used without prefolding. Thus DHX36 actually faces a linear G-rich sequence, or a partially folded structure that is not the well-folded physiological G4. Then it is possible that DHX36 could drive the non-prefolded G-rich DNA substrate into an alternatively predisposed conformation with one base shifted.

According to the model proposed by Chen *et al.*[1], G4 binding induces rearrangements of the helicase core of DHX36, through which the initial stable and canonical G4 structure (G4$^{Myc}$) are transformed into an unstable one with a non-canonical quartet and two canonical G-quartets (termed as G4$_{2G\text{-}q}$) (**Extended Data Fig. 1**). Consequently, the two guanine bases G13 and G17 originally "hidden" in the G-quartets would be translocated into a double-chain-reversal loop and a 3' single-stranded region, respectively. To see if this scenario really occurs in solution, we resort to two typical assays that are widely used to detect the stable formation of G4 structure as well as the spatial positions of G bases. First, a polymerase extension assay[2] was performed with the same ssDNA templates, but harboring different G4-forming sequences in front of the primer extension direction. We observed the polymerase can replicate the G4$_{2G\text{-}q}$ but not the G4$^{Myc}$ sequences due to the different stabilities of G4 structures they form (**Fig. 1a**). This assay thus allows us to probe the potential formation of $G4_{2G\text{-}q}$ from $G4^{Myc}$ upon DHX36 binding. We found that, in the presence of high-concentration DHX36 and under different experimental conditions, the $G4_{2G\text{-}q}$ sequence can still be replicated, whereas the $G4^{Myc}$ sequence remains non-replicatable and the primer extension is always stalled at the first guanine (**Fig. 1a**). This indicates that the binding of DHX36 does not transform the initial stable $G4^{Myc}$ into an unstable $G4_{2G\text{-}q}$. Next, since no protected N7-Guanine could be methylated by dimethyl sulfate (DMS), DMS footprinting assay was used to detect whether bases G13 and G17 are translocated away from the G-quartets into the positions as the authors indicated. As shown in **Fig. 1b**, both G13 and G17 are protected from DMS methylation, indicating that, in sharp contrast to their model, the DHX36-bound $G4^{Myc}$ keeps its initial stable G4 conformation without one-base translocation. In fact, the above results are in accordance with our recently solved crystal structure of wild-type bovine DHX36 in complex with $DNA^{Myc}$, in which the G4 structure maintains its initial three G-quartets without significant conformational rearrangement, except the DHX36-$DNA^{Myc}$ complex adopts a dimeric structure (the details will be published elsewhere).

Chen *et al.*[1] observed that in the absence of ATP, DHX36 shifts the FRET efficiency of fluorescently labelled $DNA^{Myc}$ from an initial high-FRET state 1 ($E_{\mathrm{FRET}}$ = ~0.85, corresponding to the free and well-folded $G4^{Myc}$) to oscillation between medium state 2 ($E_{\mathrm{FRET}}$ = ~0.65) and low state 3 ($E_{\mathrm{FRET}}$ = ~0.40). They interpreted the oscillation between the last two states as ATP-independent and repetitive unfolding between the DHX36-bound canonical $G4^{Myc}$ (state 2) and non-canonical $G4_{2G\text{-}q}$ (state 3). They further suggested that ATP is likely to be required only for release of DNA from the helicase. However, we and other laboratories[3–6] have found that ATP is absolutely required for DHX36-mediated G4 unfolding. To reconcile the contradictions, we used a stopped-flow assay to characterize extensively the DHX36-mediated G4 unfolding with a series of G4 sequences which are different in loop lengths and base compositions, but all fold to parallel G4 conformations. In the presence of 1 mM ATP, while G4 structures with moderate stability ($G4^{Tel}$, $G4^{Bcl\text{-}2}$ and $G4^{H35}$) are efficiently unfolded by DHX36, that with high thermostability ($G4^{Myc}$ and $G4^{3T}$) are poorly unfolded (**Extended Data Fig. 2a**), consisting with previous observations that the G4 unfolding activity of a helicase is highly sensitive to the stability of G4[6, 7]. This explains why we rarely observed G4 unfolding activity of DHX36 in our own smFRET studies with different DNA constructs that all harbor the $G4^{Myc}$ or $G4^{3T}$ structure.

In the absence of ATP or presence of the non-hydrolysable ATP analogue ATPγS, on the other hand, none of these G4 structures can be unfolded (**Extended Data Fig. 2a**). This result is critical for understanding the ATP-independent oscillation observed by Chen *et al.*[1]. Does it really result from a repetitive unfolding? In fact, they observed that the FRET efficiency recorded with their structure-guided mutants can still go to the state 3 that they supposed to be corresponding to remodeled G4 with one nucleotide translocated. Evidently, these data are in striking contrast to their own expectation and interpretation that those mutated G4-interacting residues in the DSM are implicated in G4 unfolding.

We think the logical interpretation for the observed FRET oscillation should be as follows. In the DHX36-DNA$^{Myc}$ complex, the 3' ssDNA extension and the intact G4$^{Myc}$ are bound by DHX36 via, respectively, its helicase core and G4-binding motifs (including the DSM) (**Fig. 2a**). The latter interactions are unstable, i.e., the G4$^{Myc}$ may make frequent transitions between bound and unbound states, thus giving rise to the observed FRET oscillation in the experiments. This also explains why, when the G4-interacting residues in the DSM are mutated, the FRET signal no longer oscillates but is only fixed at the low level[1], simply because G4$^{Myc}$ remains in the unbound state.

In fact, in our smFRET assay performed under similar conditions as used by Chen *et al.*[1], two types of ATP-independent FRET oscillation were observed. In a small portion of the data curves (~20%) and in accordance with previous observation by Chen *et al.*[1], the FRET signal, upon addition of DHX36, first drops from state 1 to 2 and then starts to oscillate between states 2 and 3 (**Fig. 2b**, upper panel); whereas in other FRET curves (~80%), the FRET signal first drops from state 1 to 3 directly and then starts to oscillate between states 3 and 2 (**Fig. 2b**, lower panel). According to our model, the former case corresponds to that the ssDNA extension and G4$^{Myc}$ of the DNA substrate are simultaneously bound by DHX36 at first (resulting in FRET change from state 1 to 2) and then the DHX36/G4 interactions start to fluctuate, while the latter case corresponds to that only the ssDNA extension is bound by DHX36 at first (resulting in FRET change from state 1 to 3) and then the G4$^{Myc}$ starts to be bound by DSM (resulting in FRET change from state 3 to 2). Note that the latter case cannot be explained by the model of Chen *et al.*[1].

To provide more evidences that the above observed oscillations correspond to repetitive binding, but not G4 unfolding, we recorded FRET traces with increasing concentration of DHX36. The reasoning is that, if the oscillations result from repetitive unfolding as proposed by Chen *et al.*[1], the unfolding activity will not be affected by a high concentration of DHX36. Even if it is impeded, the FRET signal would remain in state 2. However, if the oscillation arise from DSM binding as we proposed, the transient binding-dissociation between the DSM and G4$^{Myc}$ of an immobilized DNA substrate will offer the G4$^{Myc}$ an opportunity to be bound by an excess of inter-molecular DSM motifs at high concentrations of DHX36, thus the FRET oscillation will disappear and the FRET signal remain in state 3 because the intramolecular DSM is excluded to bind the G4$^{Myc}$. The results shown in **Fig. 2c** confirm our expectations.

Next, in order to further confirm our proposed model, we labeled DHX36 directly at the DSM with the dye molecule Cy5. In this case, if DHX36 really binds to G4 repetitively, the FRET signal would oscillate. As expected, this was indeed observed in the experiment (**Fig. 2d**). According to the model proposed by Chen *et al.*[1], the DSM would always remain bound to the G4, and thus the distance between Cy3 and Cy5, and the FRET signal should be constant.

Finally, we used the smFRET assay to study DHX36-mediated G4 unfolding. Cy3 was labelled adjacent to the 3' end of G4, allowing direct observation of G4 unfolding without the interference from ssDNA binding signal. We first confirmed with the stopped-flow assay that such labeled Cy3 does not block DHX36 from unfolding the G4 structure ($G4^{tel}$ was used) (**Extended Data Fig. 2b**). Then in the smFRET assay, we found that while addition of ATPγS does not change the FRET level, addition of ATP gives rise to 3–4 FRET oscillation cycles, ended with the final escape of the G4-harboring ssDNA sequence due to duplex unwinding, indicating an ATP-dependent G4 unfolding (**Fig. 2e**, left). Moreover, while no FRET variation was observed with the stable $G4^{Myc}$ upon ATP addition, unfolding signal was recorded with the less stable $G4_{2G\text{-}q}$ under the same experimental conditions (**Fig. 2e**, right). Note that the FRET traces from ATP hydrolysis-driven gradual G4 unfolding (**Fig. 2e**) and ATP-independent repetitive G4 binding (**Fig. 2b**) are radically different.

In summary, our results from polymerase extension, DMS footprinting, stopped-flow, and smFRET assays do not support the ATP-independent one-base translocation model proposed by by Chen *et al.*[1]. We believe the previously observed FRET oscillation should correspond to a repetitive G4 binding, but not unfolding by DHX36. The exact mechanism of DHX36 family helicase-mediated G4 unfolding still remains elusive.

**Methods**

Polymerase extension assay was performed essentially according to Teng *et al.*[2] with minor modifications. Briefly, DNA template (1 μM) harboring a G4 sequence was hybridized with 1 μM FAM-labeled DNA primer. The resulting DNA was folded in the presence of 100 mM KCl according to the classical G4 DNA preparation protocol. Then it was incubated with or without DHX36 in the standard reaction mixture for 30 min at 25 ℃. The polymerase extension reaction was initiated by adding 1 mM NTP and Klenow fragment (KF). After 20 min, the reaction was terminated by the addition of 2× loading buffer (8 M urea, 0.05% Xylene Cyanole) and then heated for 20 min at 98 ℃. The samples thus obtained with different DHX36 and KF concentrations were then subjected to urea (8 M) polyacrylamide (15%) gel electrophoresis, and the denatured products were visualized by a gel imaging analysis system.

For DMS footprinting assay, 3' FAM-labelled DNA was firstly folded in the presence of 100 mM KCl and then incubated with or without DHX36 for 30 min at room temperature. Then 0.1% DMS was added. After 15 min, the DNA was ethanol precipitated and cleaved with 5% piperidine at 90 ℃ for 30 min. Samples were resolved by 15% denaturing polyacrylamide gel electrophoresis.

The fluorescence-based stopped-flow and smFRET assays were performed as described in reference 8. All DNA substrates used in the study are listed in **Extended Data Table 1**.

## Author affiliations

*College of Life Sciences, Northwest A&F University, Yangling, 712100, China*
Hai-Lei Guo, Wei-Fei Chen, Na-Nv Liu, Ze-Yu Song, Xi-Miao Hou & Xu-Guang Xi

*The University Lyon, ENS de Lyon, University Claude Bernard, CNRS UMR 5239, INSERM U1210, LBMC, 46 Allée d'Italie Site Jacques Monod, F-69007, Lyon, France*
Stephane Rety

*Institute of Physics, ChineseAcademy of Sciences, Beijing, 100190, China*
Shuo-Xing Dou

*LBPA, ENS de Cachan, CNRS, Université Paris-Saclay, Cachan, F-94235, France.*
Xu-Guang Xi

**Data availability** Data are available from the corresponding author upon request.

**Author Contributions** H.L.G., W.F.C., S.R., N.N.L. Z.Y.S., X.M.H. and S.X.D performed the experiments and/or analyzed the data; W.F.C., S.R., S.X.D. and X.G.X. wrote the manuscript; X.G.X. and S.X.D. conceived the study.

**Competing interests** Declared none.

**Corresponding author** Correspondence to Xu-Guang Xi.

## Figure legends

**Fig. 1: DHX36 binding does not alter the conformation of the canonical G4 with one base translocation**.

**a**, Analysis with polymerase extension assay. Guanines in the DNA templates (PEDNA$_{2G\text{-}q}$ or PEDNA$^{Myc}$) to form G4 structures (G4$_{2G\text{-}q}$ or G4$^{Myc}$) are colored in red, FAM-labeled DNA primer is in dark gray, bases (from +1 to +26) to be added to the primer by Klenow fragment (KF) are in light gray. The only difference between the two templates is that the 3'-most guanine (bold) in PEDNA$^{Myc}$ is replaced by a thymine (blue) in PEDNA$_{2G\text{-}q}$. **b**, DMS footprinting analysis. Guanines in the ssDNA substrate (DMSDNA$^{Myc}$) to form the canonical G4 structure (G4$^{Myc}$) are colored in red. The two Guanines that would be translocated away from the G-quartets in the binding-induced non-canonical G4 structure (G4$_{2G\text{-}q}$) are in bold. The two guanines in the ssDNA substrate for control (DMSCtrl) are colored in red.

**Fig. 2: DHX36 displays ATP-independent repetitive binding to G4 and ATP-dependent G4 unfolding**.

**a**, Proposed model for ATP-independent repetitive binding of DHX36 to G4. The average distance between Cy3 and Cy5, due to thermal motion of the flexible ssDNA extension, is the shortest when the DNA substrate is not bound by DHX36 (left panel). It becomes longer when DHX36 binds both the ssDNA extension and the G4 (middle panel). It is the longest when DHX36 binds the ssDNA extension alone (right panel). **b**, smFRET assay similar to that by Chen *et al.*[1], using DNA construct formed with smDNA$^{Myc\text{-}1}$ and smbiotin$^{-1}$. Upon addition (indicated by gray arrow) of 30 nM DHX36, the FRET first drops from high (~0.9, state 1) to medium (~0.65, state 2) level (red arrow, upper panel), or first drops from high (~0.9, state 1) to low (~0.4, state 3) level (green arrow, lower panel), and then starts to oscillate between medium and low levels. The probabilities of occurrence for the two cases are ~80% and ~20%, respectively. **c**, ATP-independent FRET traces at different concentrations of DHX36. **d**, smFRET observation of ATP-independent repetitive binding DHX36 to G4$^{Myc}$. The N-terminal end of DSM of DHX36 was labeled with Cy5. The DNA construct was formed with smDNA$^{Myc\text{-}2}$ and smbiotin$^{-2}$, and 2 nM labelled DHX36 was used. **e**, Direct observation of G4 unfolding by DHX36. Fluorescent signals of Cy3 and Cy5 from unfolding of G4$^{Tel}$ in a DNA construct formed with smDNA$^{Tel}$ and smbiotin, in the presence of ATP (upper left), and the FRET traces in the presence of ATP or ATPγS (lower left). DHX36 can unfold G4$_{2G\text{-}q}$ (DNA construct with smDNA$_{2G\text{-}q}$ and smbiotin) but not G4$^{Myc}$ (DNA construct with smDNA$^{Myc}$ and smbiotin) under the same experimental conditions (right panel). 30 nM DHX36 and 1 mM ATP or ATPγS were used.

**Extended Data Fig. 1: Schematic of the G4 structures formed by DNA*$^{Myc}$*.**

**a**, Canonical G4 structure (G4$^{Myc}$) formed by the G4-forming sequence in DNA$^{Myc}$. **b**, Non-canonical G4 structure ($G4_{2G\text{-}q}$) induced by DHX36 binding. Compared with G4$^{Myc}$, bases A10, G11, G12, G13, T14, G15, G16, and G17 are translocated by one base in the 3' direction, thus bases G13 and G17 come out of the G-quartets. The non-canonical quartet (top) is composed of G2, G6, A10, and T14.

**Extended Data Fig. 2: Characterization of the G4 unfolding activity of DHX36 with different substrates.**

**a**, ATP is required for G4 unfolding. The stopped-flow assay was performed with 20 nM DHX36 helicase, 4 nM DNA substrate in the absence or presence of 1 mM ATP or ATPγS. The DNA substrates were constructed with the G4-harboring DNA sequences (SFDNA$^{Tel}$, SFDNA$^{H35}$, SFDNA$^{Bcl\text{-}2}$, SFDNA$^{3T}$, or SFDNA$^{Myc}$) and the complementary ssDNA (SFComplem). In the absence or presence of 1 mM ATPγS, no unfolding was observed for all the G4 structures. **b**, Stopped-flow assay for monitoring the unfolding of G4$^{Tel}$ in DNA substrates to be used in smFRET assay, with Cy3 labelled at the end of the 3' ssDNA tail or third base from the ssDNA/G4 junction. The assay was performed with 20 nM DHX36 helicase, 4 nM DNA substrate formed with smDNA$^{Tel}$ (or smDNA$^{Tel}$-1) and smbiotin, and 1 mM ATP.

## Extended Data Table 1 Sequences of DNA substrates used in different assays

| Assay | Substrate | Sequence (5'–3')[a] |
|---|---|---|
| Polymerase Extension | PEDNA$^{Myc}$ | A**GGG**T**GGG**TA**GGG**T**GGG**TTTTTTTTT*GTACATCAAATC* |
| | PEDNA$_{2G\text{-}q}$ | A**GGG**T**GGG**TA**GGG**T**GGT**TTTTTTTTT*GTACATCAAATC* |
| | PEPrimer | (F)*GATTTGATGTAC*[b] |
| DMS | DMSDNA$^{Myc}$ | A**GGG**T**GGG**TA**GGG**T**GGG**TTTTTTT(F) |
| | DMSCtrl | TTTTTTTTTTGTTTTTGTTTTTTTT(F) |
| Stopped-Flow | SFDNA$^{3T}$ | (H)*GGAAGGAACTGTATGTA***GGG**T**GGG**T**GGG**T**GGG**TGTTGTT[c] |
| | SFDNA$^{Tel}$ | (H)*GGAAGGAACTGTATGTA***GGG**TTA**GGG**TTA**GGG**TTA**GGG**TGTTGTT |
| | SFDNA$^{Myc}$ | (H)*GGAAGGAACTGTATGTA***GGG**T**GGG**TA**GGG**T**GGG**TGTTGTT |
| | SFDNA$^{Bcl\text{-}2}$ | (H)*GGAAGGAACTGTATGTA***GGG**CGC**GGG**AGGAAGG**GGG**C**GGG**TGTTGTT |
| | SFDNA$^{H35}$ | (H)*GGAAGGAACTGTATGTA***GGG**TGT**GGG**TGTGT**GGG**TGTGGTGTGT**GGG**TGTTGTT |
| | SFComplem | *TACATACAGTTCCTTCC*(F) |
| smFRET | smDNA$^{Tel}$ | *TGGCACGTCGAGCAGAGTT***GGG**TTA**GGG**TTA**GGG**TTA**GGG**TG/iCy3dT/TGTTGTTGTTGT |
| | smDNA$^{Tel}$-1 | *TGGCACGTCGAGCAGAGTT***GGG**TTA**GGG**TTA**GGG**TTA**GGG**TGTTGTTGTTGTTGT(Cy3) |
| | smDNA$^{Myc}$ | *TGGCACGTCGAGCAGAGTT***GGG**T**GGG**TA**GGG**T**GGG**TG/iCy3dT/TGTTGTTGTTGT |
| | smDNA$_{2G\text{-}q}$ | *TGGCACGTCGAGCAGAGTT***GGG**T**GGG**TA**GGG**T**GGT**TG/iCy3dT/TGTTGTTGTTGT |
| | smbiotin | *ACTC*/iCy5*dT/GCTCGACGTGCCA*-biotin |
| | smDNA$^{Myc}$-1 | *TGGCACGTCGAGCAGAGTT***GGG**T**GGG**TA**GGG**T**GGG**TGTGTGGTT(Cy3) |
| | smbiotin-1 | (Cy5)*ACTCTGCTCGACGTGCCA*-biotin |
| | smDNA$^{Myc}$-2 | *TGGCACGTCGAGCAGAGTT***GGG**T**GGG**TA**GGG**T**GGG**TGTGTGGTT |
| | smbiotin-2 | (Cy3)*ACTCTGCTCGACGTGCCA*-biotin |

[a] The G4-forming guanines are shown in red. The complementary sequences are in italics.

[b] F, FAM.

[c] H, hexachlorofluorescein.

**Fig. 1**

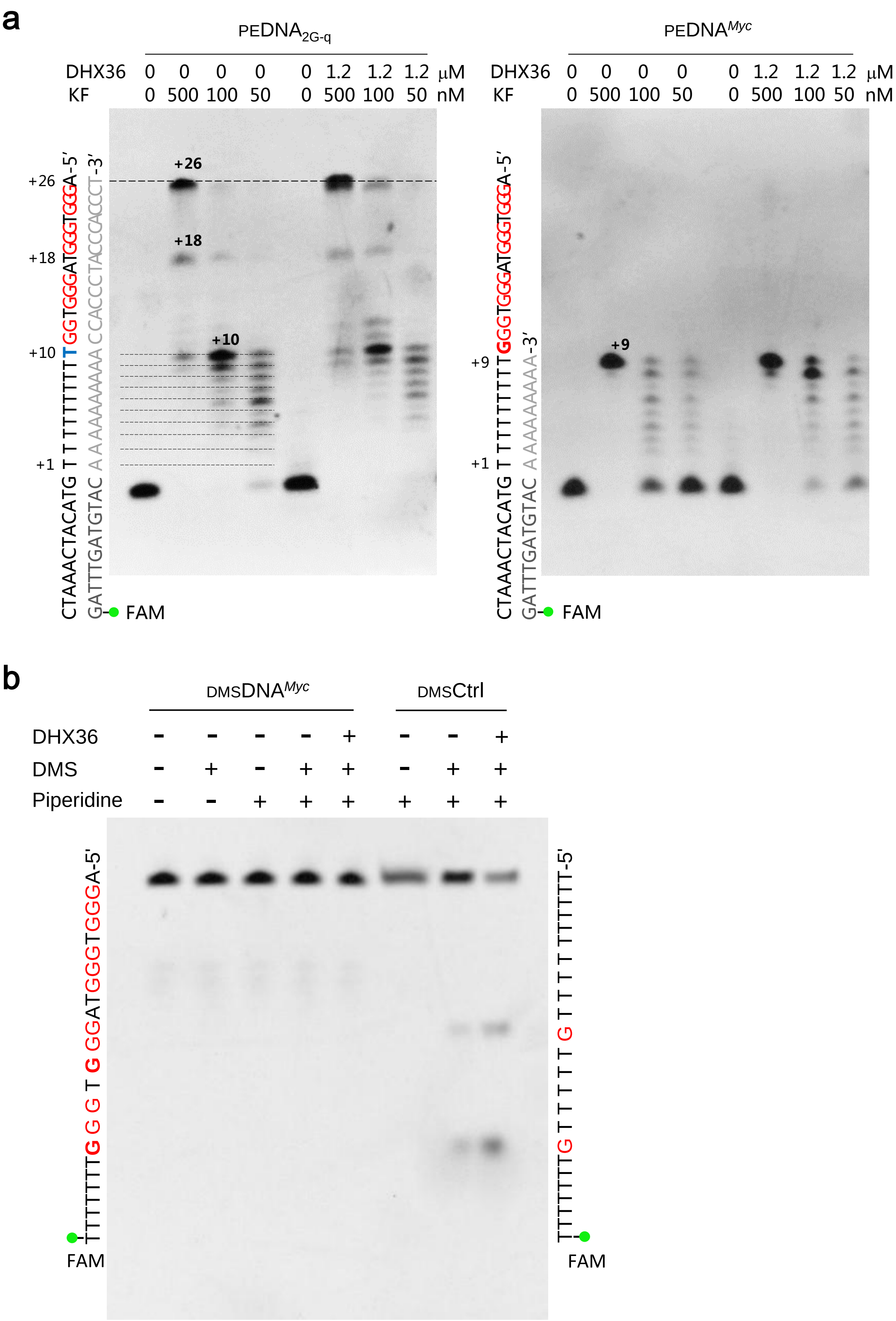
a
PEDNA2G-q
PEDNAMyc
DHX36 0 0 0 0 0 1.2 1.2 1.2 μM
KF 0 500 100 50 0 500 100 50 nM
+26
+18
+10
+9
+1
FAM
b
DMSDNAMyc
DMSCtrl
DHX36
DMS
Piperidine
FAM

**Fig. 2**

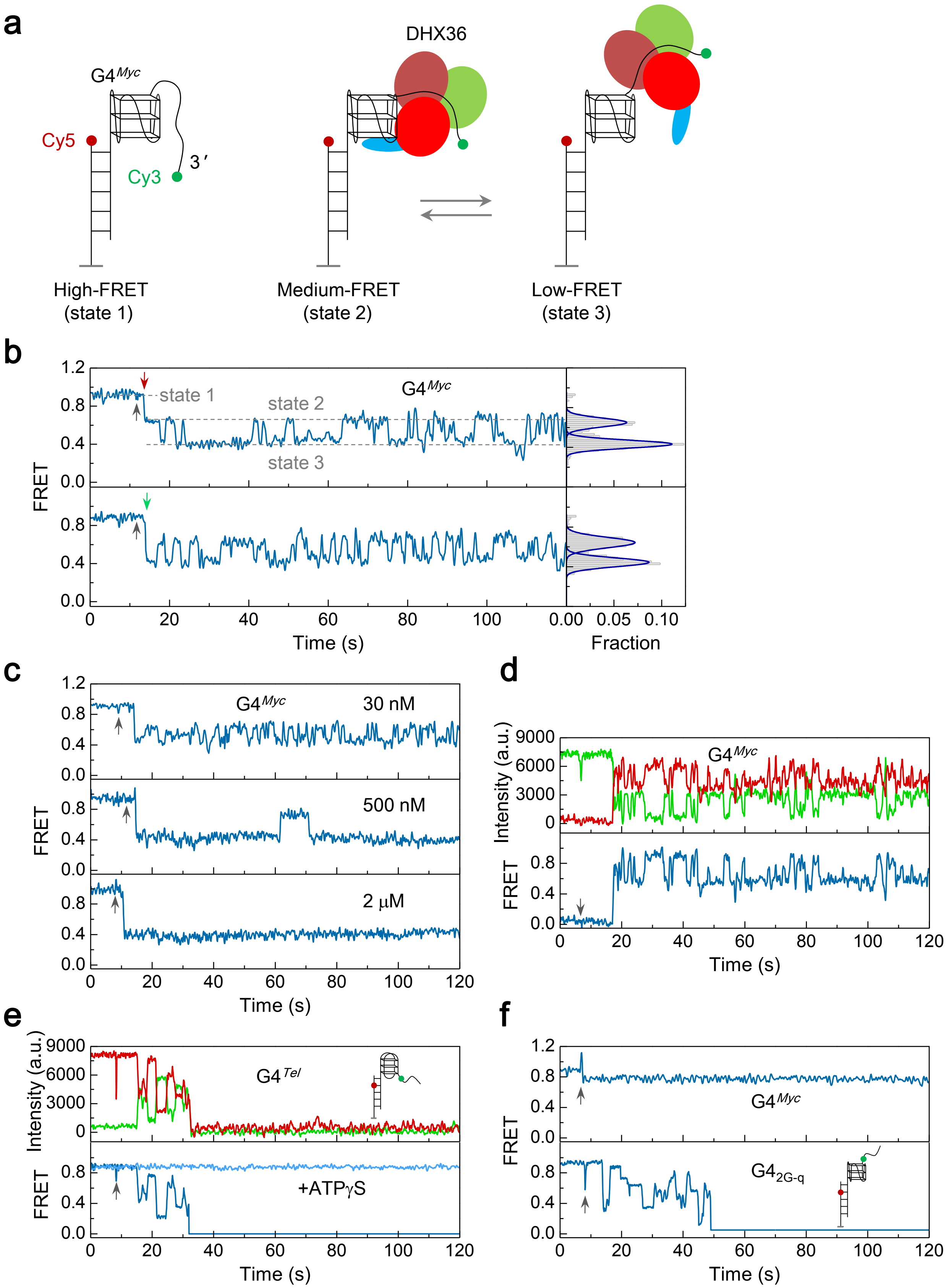

**Extended Data Fig. 1**

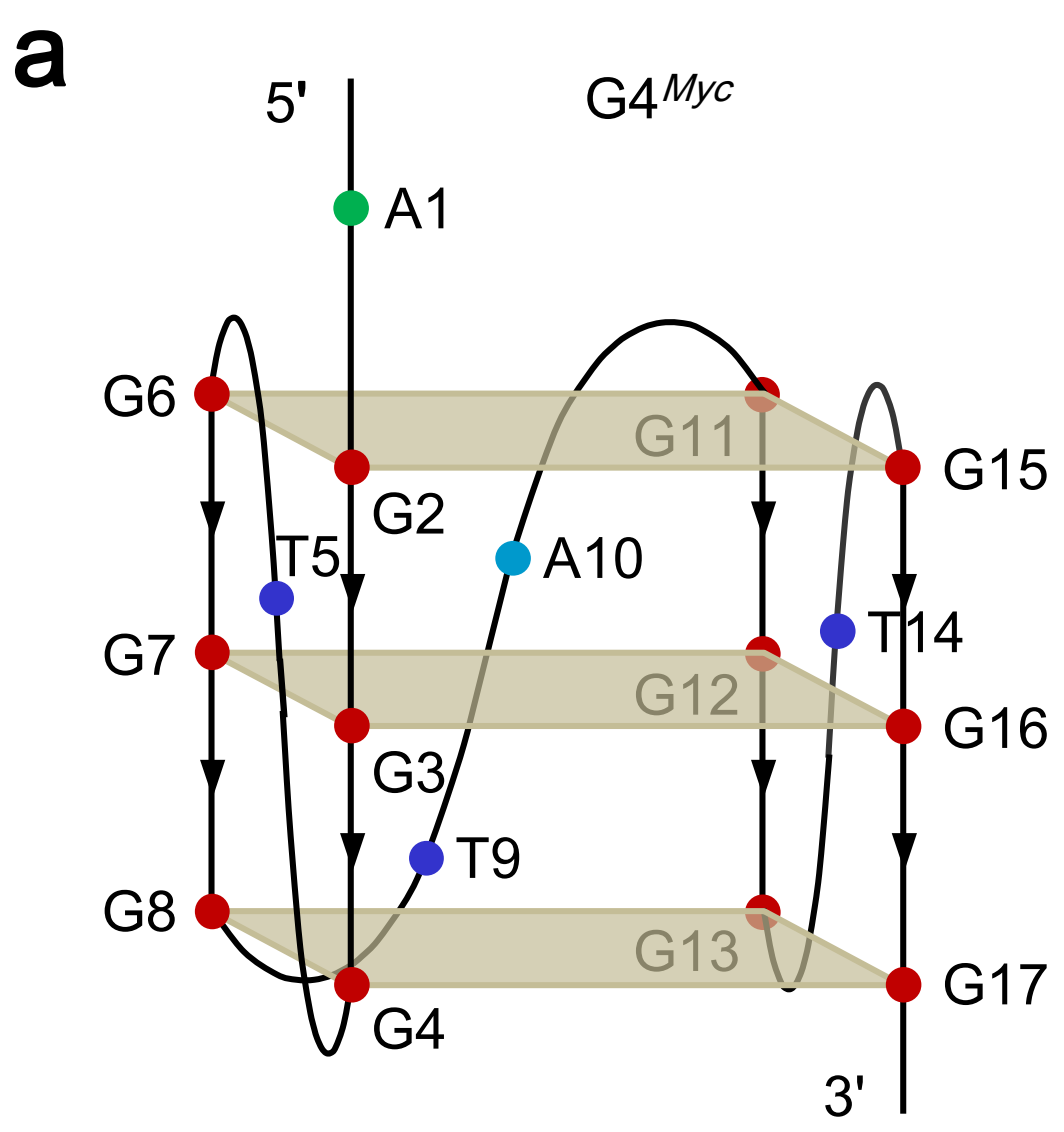

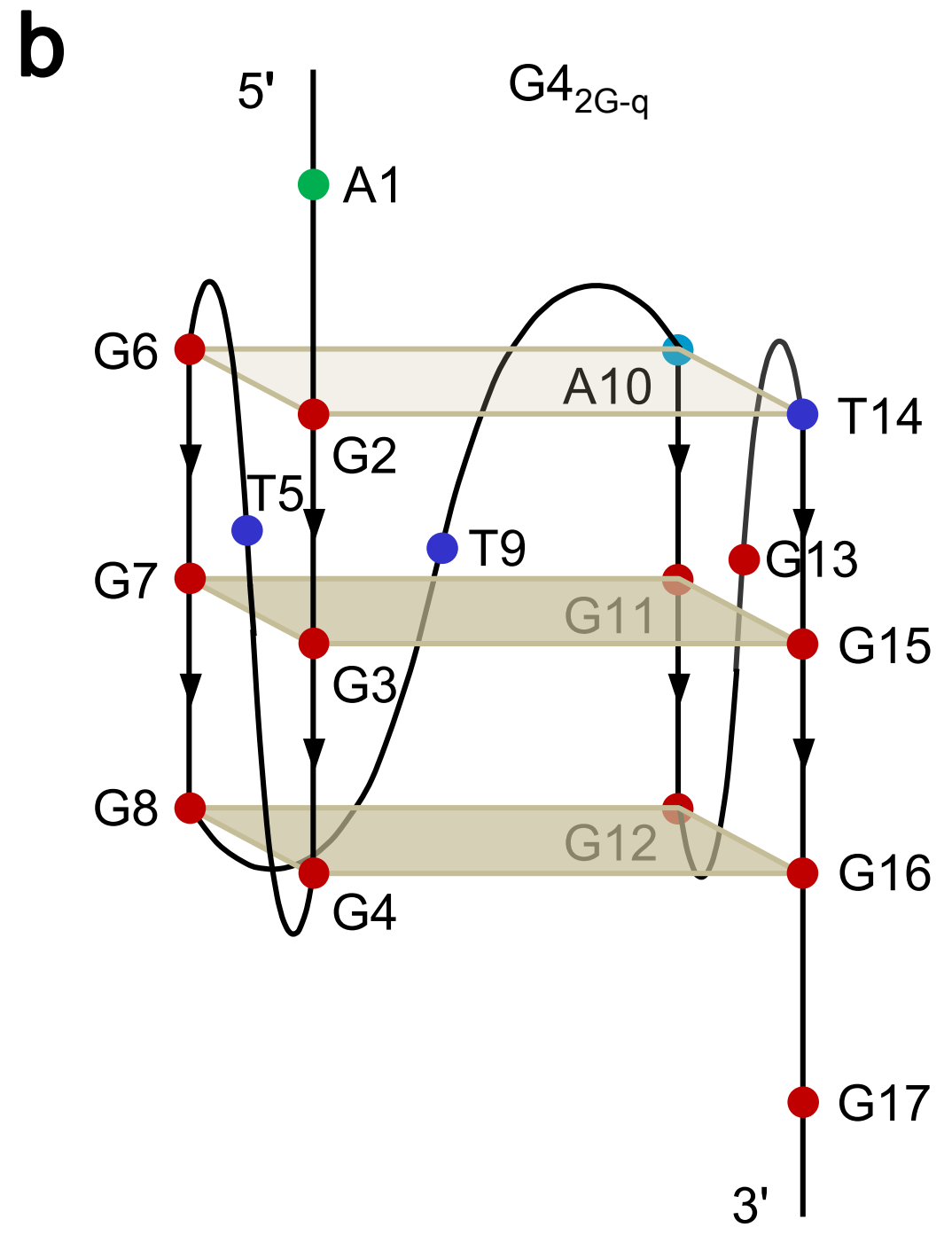

**Extended Data Fig. 2**

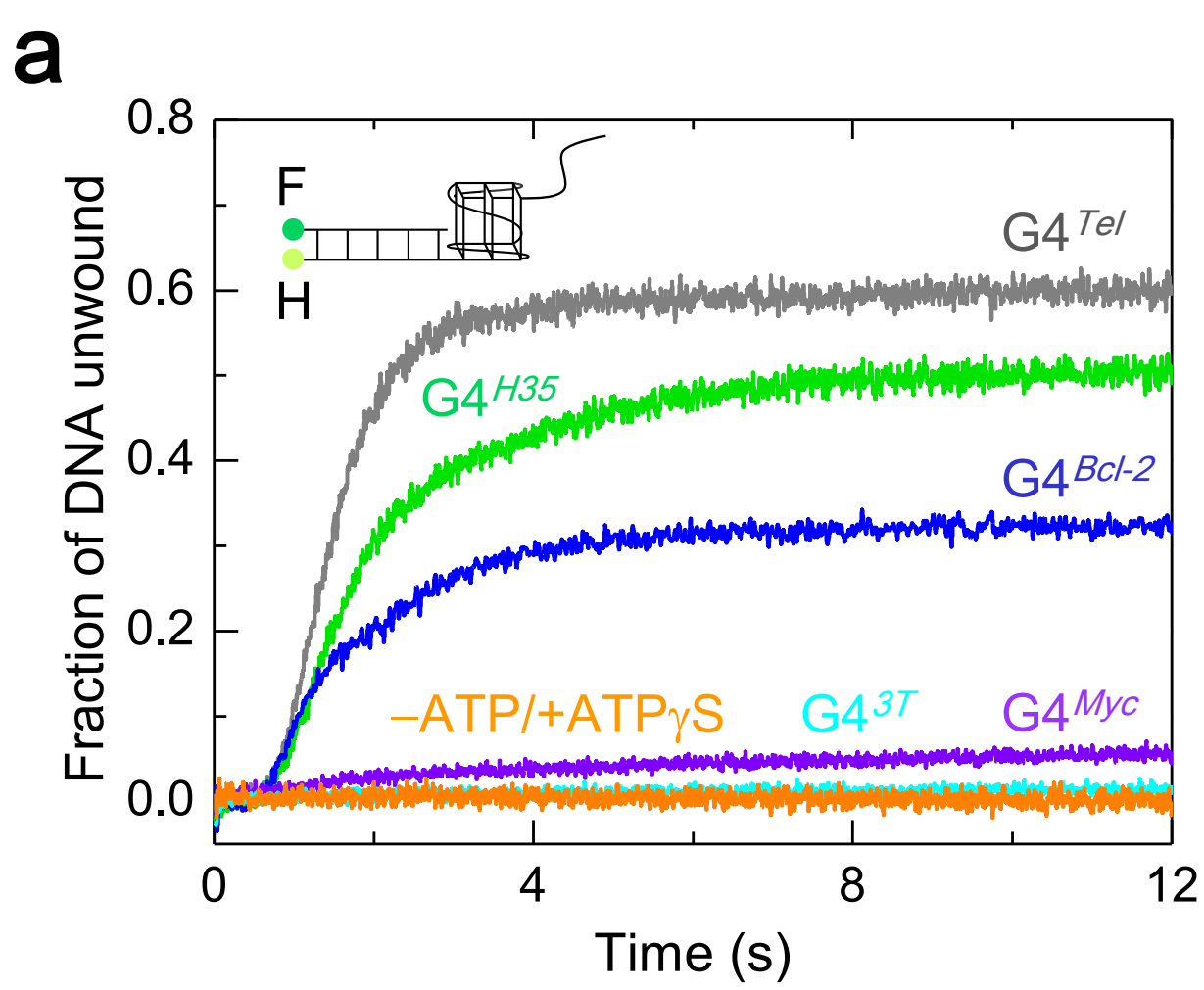

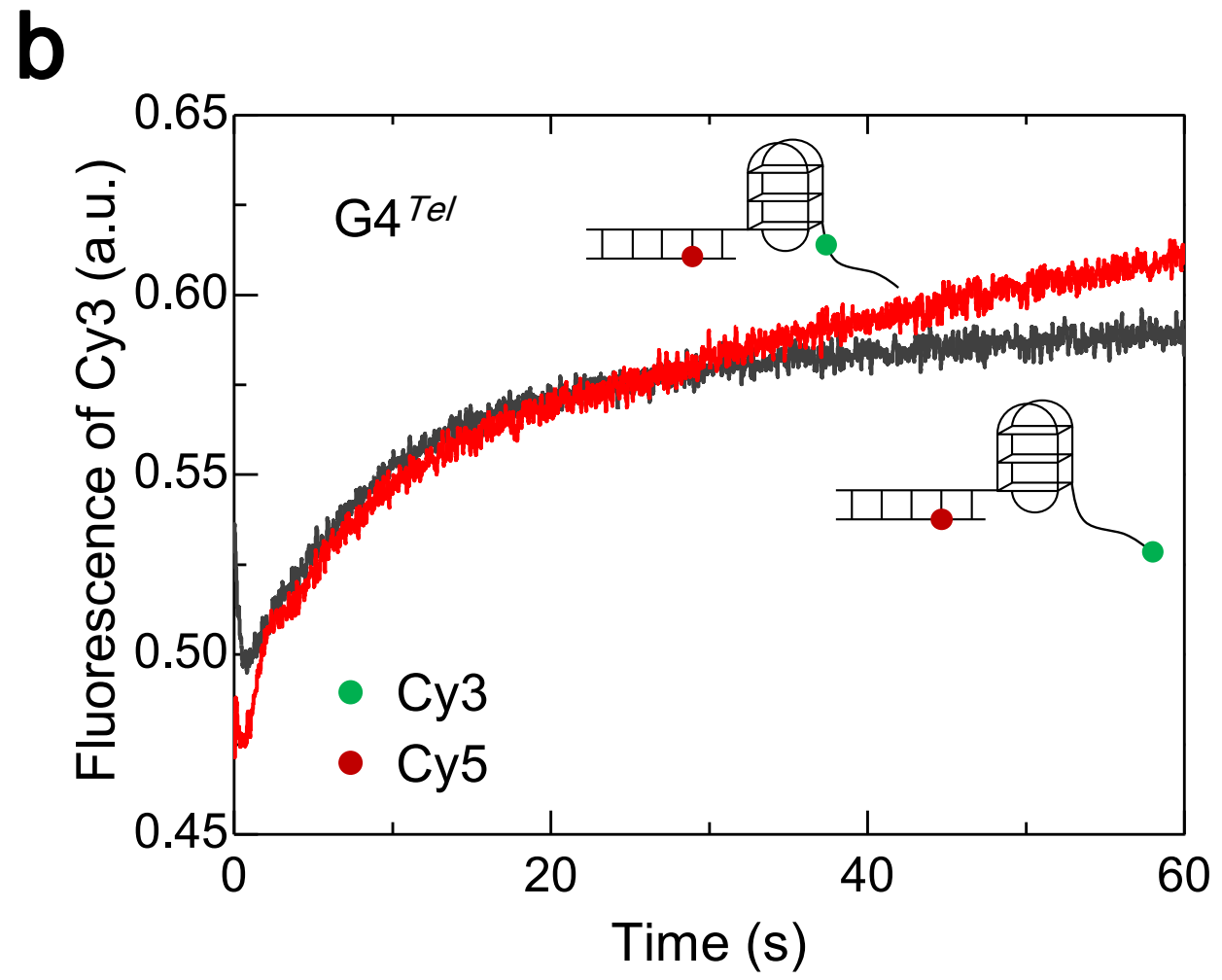